\documentclass[aps,prl,reprint,groupedaddress]{revtex4-1}

\usepackage{graphicx,amsmath,color}

\newcommand{\be}{\begin{equation}}
\newcommand{\ee}{\end{equation}}

\begin{document}

\title{Resonant tunneling and intrinsic bistability in twisted graphene structures}

\author{J. F. Rodriguez-Nieva$^{1}$, M. S. Dresselhaus$^{1,2}$, L. S. Levitov$^{1}$}
\affiliation{$^{1}$Department of Physics, Massachusetts Institute of Technology, Cambridge, MA 02139, USA}
\affiliation{$^{2}$Department of Electrical Engineering and Computer Science, Massachusetts Institute of Technology, Cambridge, MA 02139, USA}

\begin{abstract}

We predict that vertical transport in heterostructures formed by twisted graphene layers can exhibit a unique bistability mechanism. Intrinsically bistable $I$-$V$ characteristics arise from resonant tunneling and interlayer charge coupling, enabling multiple stable states in the sequential tunneling regime. We consider a simple trilayer architecture, with the outer layers acting as the source and drain and the middle layer floating. Under bias, the middle layer can be either resonant or non-resonant with the source and drain layers. The bistability is controlled by geometric device parameters easily tunable in experiments. The nanoscale architecture can enable uniquely fast switching times.
\end{abstract}


\maketitle

\section{I. INTRODUCTION}

Nanoscale systems that can switch between distinct macroscopic states upon variation of some control parameter are in high demand in diverse areas of nanoscience research. Bistable electronic systems which exhibit fast switching are of interest for applications such as low-power memory and logic \cite{memoryreview}. Recently, new realizations of intrinsically bistable system have been discovered, both in graphene \cite{flashmemory1,flashmemory2,resistivememory1,resistivememory2,fetmemory} and in other systems \cite{memristor0,memristorreview,magnetic}. In particular, van der Waals heterostructures comprising graphene layers sandwiched between insulating hexagonal boron-nitride (hBN) layers afford electronic environments with tailored band structures and transport characteristics \cite{vdwheterostructures}. It was demonstrated that introducing a twist between adjacent graphene layers in such heterostructures can result in a resonant behavior of the tunneling current and non-monotonic $I$-$V$ characteristics \cite{twistnovoselov}. It is therefore tempting to exploit twisted graphene multilayer structures as a platform for bistable and hysteretic nanoscale systems. 

\begin{figure}[b]
\centering \includegraphics[scale=1.0]{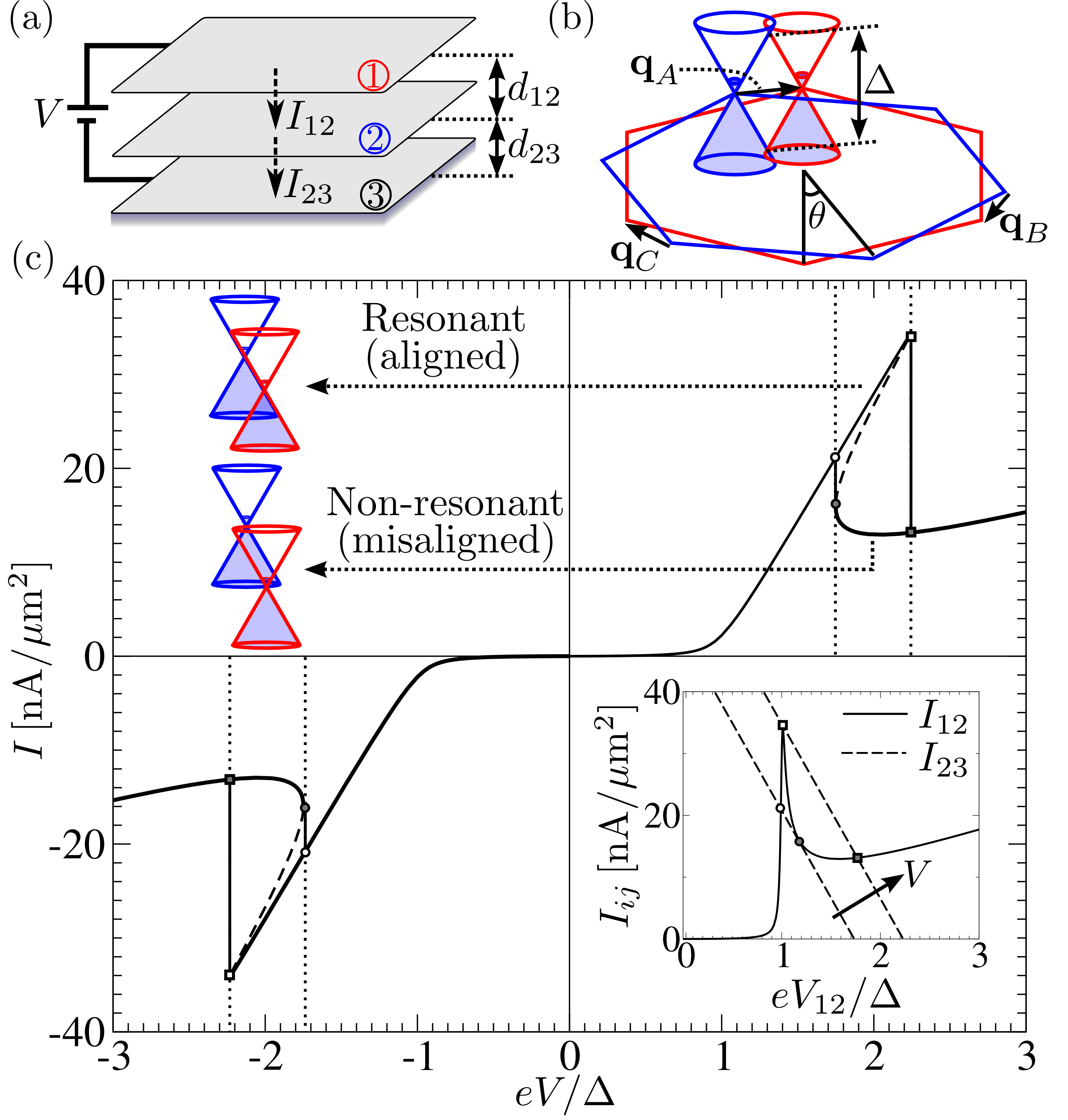}
\vspace{-4mm}
\caption{
(a) Trilayer graphene heterostructure schematics, with layers labeled 1 to 3. Here $I_{ij}$ and $d_{ij}$ are the  interlayer currents and distances. (b) Band structure of the twisted graphene layers 1(blue) and 2(red). The twist angle $\theta$ defines a characteristic energy $\Delta$ [Eq.(\ref{eq:parameters_1})] and three superlattice wave vectors ${\bf q}_{A,B,C}$ [Eq.(\ref{eq:wavevectors})]. (c) Bistable $I$-$V$ characteristics. The resonant and non-resonant bistable states are illustrated in the top left inset (details are discussed in Fig.\ref{fig:resonance}). The procedure for finding bistable solutions is illustrated in the bottom right inset [see Eq.(\ref{eq:equilibrium}) and accompanying discussion]. 
}
\vspace{-1.5mm}
\label{fig:current}
\end{figure}

Here we predict intrinsic bistability and hysteretic $I$-$V$ characteristics for vertical transport in heterostructures formed by graphene monolayers separated by hBN barriers, in a twisted arrangement similar to that described in Ref.\cite{twistnovoselov}. Essential for our bistability mechanism are resonances originating from momentum-conserving tunneling between linearly dispersing Dirac bands \cite{resonanttunneling} and occurring when the bands are aligned  \cite{twistnovoselov} (see Fig.\ref{fig:current}b,c). Bistability arises due to current-induced charge accumulation producing an interlayer bias that tunes the interband tunneling in and out of resonance. 

Below we focus on the simplest case of a two-step sequential tunneling in a device comprising three graphene monolayers. Such trilayer architecture, pictured in Fig.\ref{fig:current}a, with the top and bottom layers acting as a source and drain and the middle layer electrically decoupled (floating), is similar to  previously studied double-barrier quantum-well (QW) structures \cite{goldman}. However, our bistability mechanism, originating from resonant tunneling between Dirac bands in graphene layers, is distinct from that in the QW structures \cite{goldman}. In our case, multiple stable states arise because the decoupled layer can, for a fixed external bias, be either in a resonant (low resistance) or a non-resonant (high resistance) state. This behavior is illustrated in Fig.\ref{fig:current}c. 

The bistability is governed by geometric parameters -- the twist angle $\theta$ and the interlayer distances $d_{ij}$ -- which are easily tunable in experiments. The twist angle controls the Dirac cones' displacement in the two layers and the energy at which the cones intersect (see Fig.\ref{fig:current}b),
\be\label{eq:parameters_1}
|{\bf q}_A|= (8 \pi / 3a_0 )\,{\rm sin}(\theta/2)
,\quad
\Delta = \hbar v_{\rm F} |{\bf q}_A|
,
\ee
where $v_{\rm F}\approx 10^{6}\,{\rm m/s}$ is the carrier velocity and $a_0 \approx 2.46\,{\rm \AA}$ is the graphene lattice constant. The distances $d_{ij}$, marked in Fig.\ref{fig:current}a, determine the interlayer tunnel conductance values $G_{ij}\sim e^{2d_{ij}/\lambda}$, where $\lambda$ is the WKB length governing the  tunneling amplitude dependence on barrier width. In what follows we will use the conductance ratio
\be
Z=G_{12}/G_{23}\sim e^{2(d_{23}-d_{12})/\lambda}
\label{eq:parameters_2}
\ee
where $G_{ij}$ denotes the conductance between the corresponding layers. 

The quantities $\theta$ and $d_{ij}$ can be controlled with a large degree of precision. The twist angle $\theta$ can be tuned within $\sim 1^{\circ}$ during fabrication \cite{twistnovoselov}, whereas $d_{ij}$ can be varied by adding monolayers of dielectric materials, such as hBN or MoS$_2$. Since typical values $\lambda = \hbar / (2 m_e W)^{1/2}\sim 2\,{\rm \AA}$, estimated for the tunneling barrier height $W \sim 1\,{\rm eV}$ and the effective electron mass $m_e\sim 10^{-30}\,{\rm kg}$, are comparable to the hBN or MoS$_2$ monolayer thickness, variation in $d_{ij}$ results in a fairly gradual change of $Z$.

One appealing aspect of this system is the short interlayer transport length of a nanometer scale, which can allow high operation speeds and fast switching times. This is evident from an estimate for the $RC$ time, $\tau_{RC} = \kappa/4\pi  gd \sim 100\,{\rm ns}$, where $\kappa \sim 1$ is the dielectric constant, $d\sim 1\,{\rm nm}$ is the interlayer separation, and $g \sim 10^{-7}\,\Omega^{-1}\mu{\rm m^{-2}}$ is the interlayer conductance per unit area. The combination of geometric tunability and small transport lengths is not present in previously studied graphene-based bistable systems, such as graphene flash memories \cite{flashmemory1,flashmemory2} or graphene resistive memories \cite{resistivememory1,resistivememory2,fetmemory}. Small thicknesses can also enable large packing densities. 

The steep electronic dispersion in graphene makes the bistable state properties distinct from those in QW systems. In our case, the bistability is controlled by the resonances arising due to band alignment.
The corresponding bias value, which scales as a power law of the energy $\Delta$ given in Eq.\eqref{eq:parameters_1}, can be as large as $\delta V \sim 100$-$500\,{\rm mV}$ (see discussion below). In QW systems, instead, the bias range where bistability occurs is mainly controlled by the amount of charge $n_{\rm QW}$ that can be stored in a quantum well, $\delta V \approx en_{\rm QW}/C$, where $C$ is the interlayer capacitance \cite{sheardtombs}. Typical carrier densities in the ``charged'' and ``uncharged'' states of a bistable QW system, assessed by magnetic oscillation measurements\cite{goldman2},  are on the order of $n_{\rm QW} \sim 10^{11}/{\rm cm^2}$ and $n_{\rm QW} \sim 0$, respectively. These carrier densities yield typical values $\delta V \sim 50\,{\rm mV}$ in double-barrier quantum wells with a width of tens of nanometers ($C\sim 0.1$-$1\,{\rm mF}$). Such values can be as much as an order of magnitude smaller than the above estimate predicts for the graphene case. 

\section{II. SEQUENTIAL TUNNELING MODEL}

Vertical transport in our trilayer architecture can be described by a simple sequential model. The model validity relies on the interlayer tunnel coupling being weak such that the inter-layer charge transfer is slow compared to the intra-layer electron relaxation. Indeed, the values $\tau_{RC}$, estimated above, are much longer than typical thermalization times in graphene, $\tau_{\rm th} \sim 10\,{\rm ps}$ \cite{relaxation}. The $RC$ times, however, are sufficiently fast to be competitive with the speeds of existing switching devices \cite{memoryreview}. 

The interlayer transport mechanism is mainly governed by the twist angle $\theta$, which defines the K-point displacement  ${\bf q}_A$ between graphene lattices in adjacent layers, and the interlayer bias. 
Under bias, the value $|{\bf q}_A| $ given in Eq.\eqref{eq:parameters_1} determines the range of momenta and energies for which momentum-conserving tunneling is allowed. Large values of $|{\bf q}_A|$ hinders resonant tunneling given that phonon and defect scattering are necessary to supply the large momentum mismatch between layers. Momentum-nonconserving transport can also occur if the top/bottom layer is made of a different material so that there is a large mismatch between unit cells with respect to that of graphene. For small $|{\bf q}_A|$, on the other hand, momentum conserving tunneling is possible for moderately small values of bias. 

In our two-step sequential tunneling model, we treat transport between layers 1 and 2 as momentum-conserving. The second step, between layers 2 and 3, is assumed to be momentum-nonconserving and described by Ohm's law. The latter assumption allows us to simplify our discussion and focus on the essential aspects of bistability. In addition, we also assume that the contact resistances are sufficiently small so that all the potential drop occurs predominantly between the graphene layers. 

Turning to a systematic development of the model, the low energy Hamiltonian ${\cal H}$ describing coherent transport between a pair of twisted graphene monolayers has contributions ${\cal H}={\cal H}_1 + {\cal H}_2 + {\cal T}_{12}$. Here ${\cal H}_{1,2}$ are the free-particle terms describing massless Dirac particles in each graphene layer, and ${\cal T}_{12}$ describes the interlayer tunnel coupling \cite{twistedacn,twistedmd1,twistedmd2}. The free particle terms are 
\be
\begin{array}{c}
\displaystyle {\cal H}_1 = \displaystyle\sum_{\bf k}\psi_{1,{\bf k}}^\dagger [\hbar v_{\rm F}{\boldsymbol\sigma\cdot({\bf k}}+{\bf q}_A/2) - \mu_1]\psi_{1,{\bf k}}, \\
\displaystyle {\cal H}_2 = \sum_{\bf k}\psi_{2,{\bf k}}^\dagger [\hbar v_{\rm F}{\boldsymbol\sigma\cdot({\bf k}}-{\bf q}_A/2)-\mu_2]\psi_{2,{\bf k}},
\end{array}
\label{eq:H0}
\ee
where $\mu_{1,2}$ are the Fermi energies measured relative to the Dirac point. For a small twist angle $\theta$, the large-wavenumber processes that couple different valleys can be neglected. In this case, it is sufficient to account for a single Dirac cone in each layer, see Eq.(\ref{eq:H0}). 
We adopt this approximation below. 

\begin{figure}
\centering \includegraphics[scale=1.0]{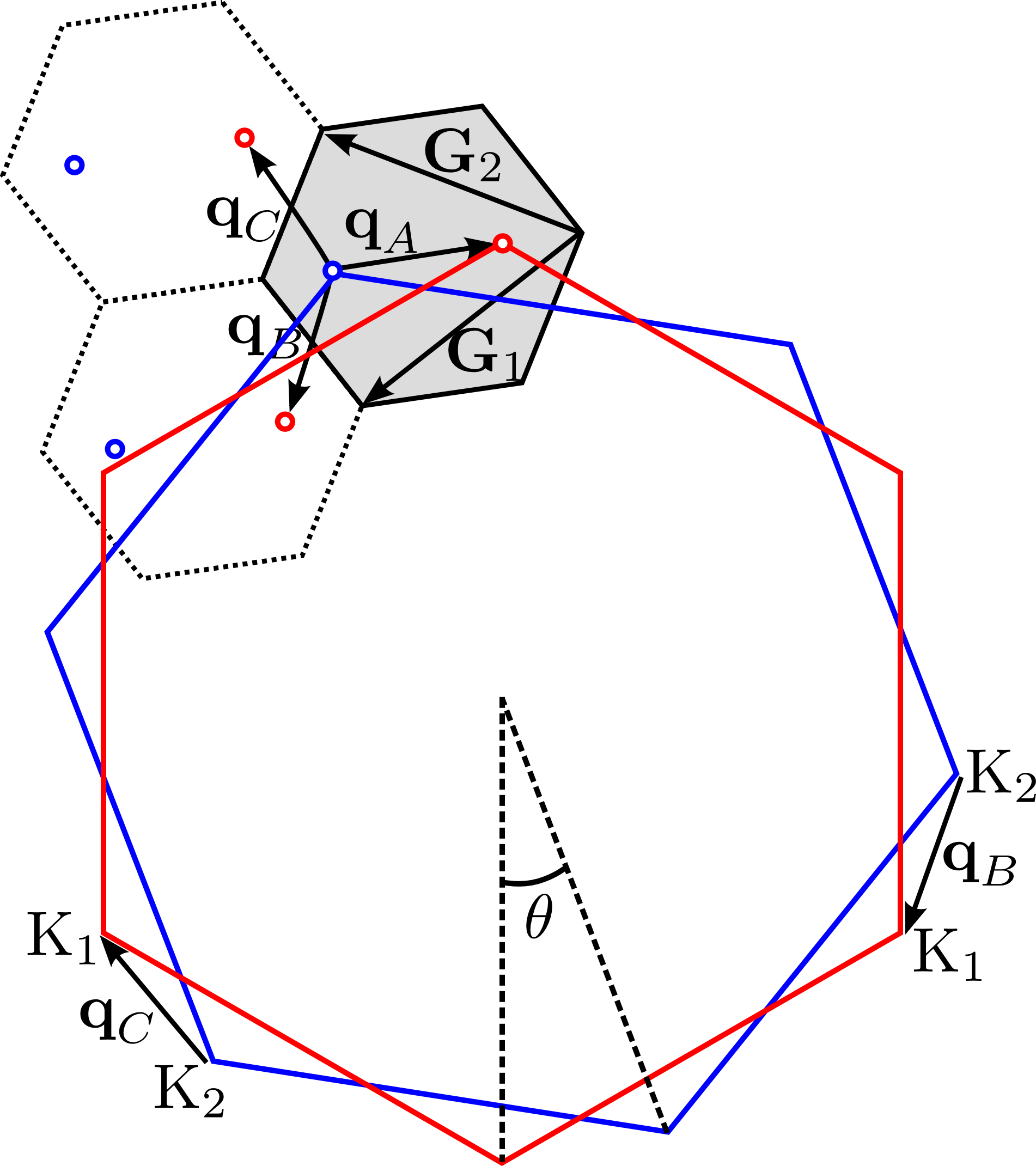}
\caption{Twisted graphene layers form an hexagonal superlattice with reciprocal superlattice vectors ${\bf G}_1$ and ${\bf G}_2$ \cite{twistedacn}. The momentum conserving tunneling coupling has the periodicity of the superlattice and can be decomposed into Fourier components ${\bf G} = n{\bf G}_{1} + m {\bf G}_{2}$, with $n,m$ being integers. For a small twist angle $\theta$, tunneling is dominated by the smallest wavevectors ${\bf q}_A$, ${\bf q}_B = {\bf q}_A - {\bf G}_{1}$ and ${\bf q}_C = {\bf q}_A - {\bf G}_2$, see Eq.(\ref{eq:H0}).
}
\label{fig:superlattice}
\end{figure}

The tunneling coupling can be modeled as a local, periodic function of position \cite{twistedacn}: 
\be
{\cal T}_{12} = \sum_{\bf k,G}\psi_{1,{\bf k}}^\dagger {\bf T}_{\bf G} \psi_{2,{\bf k + G}} + {\rm H.c.}.
\ee
The periodicity of the interlayer coupling, quantified by the ${\bf G}$ wavevectors, is determined by the hexagonal superlattice unit cell that is formed by the twisted graphene layers, see Fig.\ref{fig:superlattice}. For small $\theta$, only the longest wavelength contributions are relevant for tunneling. Referred from the Dirac point of layer 1, such long wavelength components are given by ${\bf q}_A$, ${\bf q}_B = {\bf q}_A - {\bf G}_1$ and ${\bf q}_C = {\bf q}_A - {\bf G}_2$ (see Fig.\ref{fig:superlattice}), where ${\bf G}_{1,2} $ are the reciprocal vectors of the superlattice Brillouin zone, which is smaller than the graphene Brillouin zone by a factor $\sim\sin^2(\theta)$. While the higher-${\bf q}$ harmonics of the interlayer hopping potential spatial modulation also contribute to tunneling, it can be shown that their contributions vanish rapidly on the reciprocal lattice vector ${\bf G}_{1,2} $ scale \cite{twistedacn,twistedmd1}. This leads to the tunneling Hamiltonian
\be
   {\cal T}_{12} = \sum_{j=A,B,C}\sum_{\bf k} \psi_{1,{\bf k}}^{\dagger} \, {\bf T}_j \, \psi_{2,{\bf k}+{\bf q}_j} + {\rm H.c.}
   \label{eq:Tmatrix}
\ee 
comprised of only three Fourier components. In this expression for ${\cal T}_{12}$, the ${\bf k}$ vectors are relative to the Dirac point of each layer, i.e. ${\bf k} - {\bf q}_A/2 \rightarrow {\bf k} $ in layer 1 and ${\bf k} + {\bf q}_A/2 \rightarrow {\bf k} $ in layer 2. 

Parenthetically, the lattice of the dielectric material separating the graphene layers can produce slowly varying spatial modulation of the tunneling transition amplitude ${\bf T}$ in Eq.(\ref{eq:Tmatrix}), giving rise to the effects resembling those due to a twist angle $\theta$. This would be the case when the dielectric and graphene are nearly lattice-matched as e.g. in highly-oriented hBN-graphene structures, which have a small lattice mismatch of about 1.8\% (a detailed discussion of these effects can be found in Ref.\cite{brey}). Such effects, if present, would alter the values $\mathbf{q}_{A(B,C)}$ but otherwise not change our discussion in an essential way. 

Under an interlayer bias potential $V_{12}$, the tunneling current $I_{12}$ is 
\be
\begin{array}{rr}
I_{12} = & \displaystyle \frac{ e N }{\hbar}\sum_{{\bf k}ss'j}|T_j^{ss'}({\bf k})|^2 \int_{-\infty}^{\infty} \, \frac{d\omega}{2\pi} A_{1,s}({\bf k},\omega)  \\ & \displaystyle \times A_{2,s'}({\bf k}+{\bf q}_j,\tilde{\omega}) \left[ f_1(\omega)-f_2(\tilde{\omega})\right],
\end{array}
\label{eq:current}
\ee
where $s$ ($s'$) refers to the electron ($+$) and hole ($-$) bands of layer 1 (2), and $N=4$ is the spin and valley degeneracy. The functions $f_i(\omega)=1/[e^{\beta(\omega - \mu_i)}+1]$ are the Fermi distribution functions for each layer, with $\beta =1/k_{\rm B}T$ being the inverse thermal energy and $\mu_i$ being the Fermi energies. The function $A_{i,s}$ is the spectral function of layer $i$ and band~$s$. The energy for the quantities in layer 2 is offset by $\tilde{\omega} = \omega + e\Phi_{12}$ due to the built-up interlayer electrostatic potential $\Phi_{12}$ [see Eq.(\ref{eq:H0})] between layers 1 and 2. Because of capacitance effects, the interlayer electrostatic and chemical potentials are related by 
\be
eV_{12} = \mu_1-\mu_2-e\Phi_{12}, 
\label{eq:electrostatics}
\ee
where $\mu_i$ and $\Phi_{12}$ are implicit functions of $V_{12}$. 
The quantity $T_j^{ss'}$ in Eq.(\ref{eq:current}) denotes 
\be
\begin{array}{c} 
T_{j}^{ss'}({\bf k})= \langle {\bf k},s,1 | {\bf T}_{j} | {\bf k}+{\bf q}_{j},s',2\rangle, \\
\displaystyle|{\bf k},s,i\rangle = \frac{1}{\sqrt{2}} \left(\begin{array}{c} 1  \\ s e^{i\theta_{\mathbf{k}}}\end{array}\right), 
\end{array}
\label{eq:phasefactor}
\ee
where $|{\bf k},s,i\rangle $ are the two-component eigenvectors of ${\cal H}_{1,2}$ in Eq.(\ref{eq:H0}) and $\theta_{\bf k}$ is the ${\bf k}$-vector polar angle. 

The bistability can now be described by combining relations (\ref{eq:H0}) and (\ref{eq:electrostatics}) as follows. In a steady state, there is zero net flow of carriers into the middle layer. Therefore, when the external bias $V = V_{12}+V_{23}$ between top and bottom layers is fixed, the equilibrium current $I$ is obtained by solving for $V_{12}$ from the non-linear equation
\be
I(V)=I_{12}(V_{12})=I_{23}(V-V_{12}).
\label{eq:equilibrium}
\ee
This procedure to obtain the $I$-$V$ response is shown graphically in the inset of Fig.\ref{fig:current}c. The straight line describes transport between layers 2 and 3 which is assumed to follow Ohm's law, $I_{23}=G_{23} V_{23}$, where $G_{23}$ and $V_{23}$ are the interlayer conductance and interlayer bias potential between layers 2 and 3, respectively.

\section{III. ELECTROSTATIC FEEDBACK}

In order to include the electrostatic feedback, Eq.(\ref{eq:equilibrium}) needs to be complemented with further electrostatic considerations that relate the variables $V_{ij}$, $\Phi_{ij}$, and $\mu_i$. It is important to note that all variables can be determined once the carrier densities in each layer, $n_1$, $n_2$ and $n_3$, are known. Indeed, assuming that there is no external gate, the neutrality condition relates the charge densities in the different regions of the device as 
\be
n_1+n_2+n_3=0.
\label{eq:cn}
\ee
Furthermore, the application of an external bias potential $V$ fixes the Fermi level difference between layer 1 and layer 3 as
\be
eV=\mu_1-\mu_3+\frac{4\pi e^2}{\kappa}(n_1d_{13}+n_2d_{23}).
\label{eq:biaseq}
\ee
Here $d_{ij}$ is the interlayer distance between layer $i$ and layer $j$, $\kappa$ is the dielectric constant of the barrier material, and $\mu_i = {\rm sgn}(n_i)\hbar v_{\rm F} \sqrt{\pi n_i}$. In Eq.(\ref{eq:biaseq}), we implicitly assume that all layers are undoped at $V=0$. Equations\,(\ref{eq:equilibrium})-(\ref{eq:biaseq}) then form a closed set of equations from which $n_1$, $n_2$ and $n_3$ can be obtained. The remaining variables, $V_{ij}$ and $\Phi_{ij}$, are functions of $n_i$. In particular, the electrostatic potentials are $\Phi_{12}=-4\pi e^2 d_{12} n_1 /\kappa$ and $\Phi_{23}= 4\pi e^2(n_1d_{13}+n_2d_{23})/\kappa$, whereas the interlayer bias potentials are  $V_{12}=\mu_1-(\mu_2+\Phi_{12})$, and $V_{23} = (\mu_2+\Phi_{12})-(\mu_3+\Phi_{23})$.

For simplicity, here we fix the Fermi energies in Eq.(\ref{eq:current}) to a constant value $\mu_{i} = \mu$. This is equivalent to turning off capacitance effects. In this case, $V_{ij} = \Phi_{ij}$  (see Fig.\ref{fig:resonance}). This approximation is valid in the regime $4e^2d_{ij}\Delta/\kappa(\hbar v_{\rm F})^2 \approx 15\cdot d_{ij}[{\rm nm}]\Delta[{\rm eV}]/\kappa \gg 1$. In this regime, minimal changes in carrier concentration induce large interlayer electrostatic potentials. The more realistic scenario which includes quantum capacitance effects\,\cite{qc}, such that $\mu_{1,2}$ vary with $V_{12}$, is here considered in Appendix A. However, this more realistic picture only introduces small corrections to the tunneling current without major consequences to our bistability discussion. 

\section{IV. MODEL PARAMETERS}

In order to solve Eq.(\ref{eq:equilibrium}), we need to specify the matrix elements $\mathbf{T}_{j}$ in Eq.(\ref{eq:phasefactor}). A simple and explicit model for $\mathbf{T}_{j}$ and the wavevectors ${\bf q}_j$ is provided by Ref.\cite{twistedacn}:
\be
\mathbf{T}_{j}= t \left( \begin{array}{cc} e^{i\varphi_j} & 1 \\ e^{-i\varphi_j} & e^{i\varphi_j} \end{array}\right), \quad {\bf q}_{j}=\frac{\Delta}{\hbar v_{\rm F}} (\sin\varphi_j,-\cos\varphi_j), 
\label{eq:wavevectors}
\ee
with $\varphi_A=0$, $\varphi_B=2\pi/3$, $\varphi_C=4\pi/3$. This representation is obtained for small twisting angles after performing a $\theta$ rotation of phase space in layer 2 (see details in Ref.\cite{twistedacn}). 
It is also implicit in Eq.(\ref{eq:wavevectors}) that the top and bottom graphene lattices have a common lattice point \cite{twistedacn}; a rigid horizontal translation between lattices adds an additional overall phase to the matrix $\mathbf{T}_{j}$ \cite{twistedmd1}. We stress, however, that relative phases in ${\bf T}_j$ do not alter in any significant way the physics of tunneling in Eq.(\ref{eq:current}). Furthermore, while the interlayer hopping amplitude $t$ is sensitive to several parameters, e.g. twist angle \cite{twistedmd2} and the choice of dielectric material \cite{brey}, its order of magnitude is mainly governed by the wavefunction overlap between the graphene layers. Such dependence will be described below within the WKB approximation. Equations (\ref{eq:H0}) and (\ref{eq:wavevectors}) are expected to be accurate for twist angles $\theta \lesssim 10 ^{\circ}$, and energies of $1\,{\rm eV}$\,\cite{twistedmd2}. 

For an estimate below we use the value $\theta = 2^{\circ}$. This defines an energy scale $\Delta=0.37\,{\rm eV}$. Furthermore, we take a Lorentzian spectral function in Eq.(\ref{eq:current}) for both layers, $A_{i,s}({\bf k},\omega)=2\Gamma/\left[(\omega-s\hbar v_{\rm F}|{\bf k}|)^2+\Gamma^2\right]$ with the linewidth $\Gamma\sim 10$ meV. A finite linewidth $\Gamma$ is necessary to have a finite value of the peak current when $eV_{12} = \Delta$ (see Fig.\ref{fig:resonance}). The temperature and Fermi level of the system were taken to be $T=0$ and $\mu_i = 0$, respectively. With reference to Eq.(\ref{eq:current}), we define the interlayer conductance
\be
G_{12}=Sg_{12}, \quad g_{12}=2\pi N \frac{|t|^2}{(\hbar v_{\rm F})^2} \frac{e^2}{h},
\label{eq:conductance}
\ee
where $S$ is the surface area of the device. The value of $g_{12}$ is sensitive to the twist angle and the stacked dielectric material, if any, via the parameter $t$. Here we use $g_{12} = 10^{-7}\Omega^{-1}\mu{\rm m^{-2}}$. Similar values of $g_{12}$ were measured in resonant tunneling devices which contained four layers of BN in-between the graphene layers \cite{resonanttunneling}. For $Z$, we consider a value of $Z = G_{12}/G_{23} = 0.2$.

\section{V. BISTABLE $I$-$V$ CHARACTERISTICS}

\begin{figure}
\centering \includegraphics[scale=1.0]{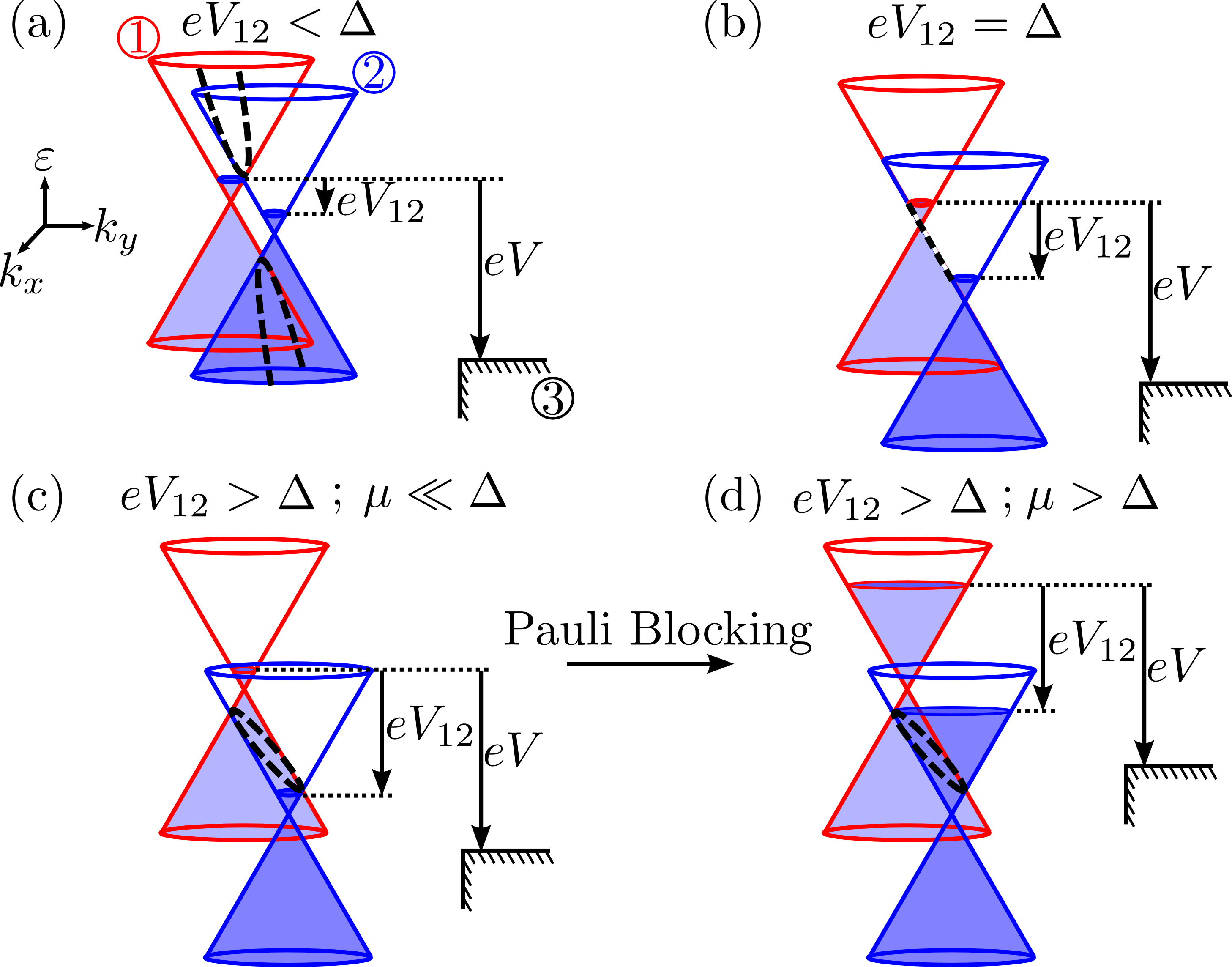}
\caption{
Regions in ${\bf k}$-space contributing to the resonant tunneling current for fixed $V$. These regions, indicated with black dashed lines at the intersection of the twisted Dirac cones, form conical surfaces in the ${\bf k}$-plane: when $eV_{12}<\Delta$ the lines form hyperbolic curves, and when $eV_{12}>\Delta$ the lines form ellipsoidal curves. When $eV_{12}=\Delta$, a van-Hove singularity in the tunneling density of states is obtained. As shown in (d), the non-resonant (high-resistance) bistable state ($eV_{12}>\Delta$) can be Pauli-blocked by adjusting the doping. Doping thus affords a way to tune the current ratio between bistable branches in Fig.\ref{fig:current}a. In this work it is assumed  that the Dirac cones are aligned at $V=0$, and that capacitance effects are neglected. Layers are labeled 1-3 as in Fig.\ref{fig:current}a. 
}
\label{fig:resonance}
\vspace{-5mm}
\end{figure}

The bistable $I$-$V$ characteristics are shown in Fig.\ref{fig:current}c. For a sufficiently large bias, $eV \gtrsim \Delta$, the current branches into two stable states. The low-resistance branch in Fig.\ref{fig:current}c corresponds to two layers at resonance (i.e., $eV_{12} \approx \Delta $), whereas the high-resistance branch corresponds to a non-resonant state (i.e., $eV_{12} > \Delta$). We note that a third solution is also possible, indicated by a dashed line in the $I$-$V$ response (see Fig.\ref{fig:current}c). This solution, however, is unstable given that a small perturbation in $\delta V_{12}$ will push the system away from this state. 

The bistable bias range can be estimated as $\delta V  \approx  (I_{12}^{\rm (pk)} - I_{12}^{\rm (vl)})/G_{23} $, where $I_{12}^{\rm (pk)}$ is the peak interlayer current and $I_{12}^{\rm (vl)}$ is the valley interlayer current (see inset of Fig.\ref{fig:current}c). To estimate $I_{12}^{\rm (pk)}$ and $I_{12}^{\rm (vl)}$, we first note that the tunneling matrix element $T_j^{ss'}({\bf k})$ varies, upon integration in ${\bf k}$-space, in the range $0 \le |T_j^{ss'}({\bf k})| \le 2t$ taking typical values $|T_j^{ss'}({\bf k})|\approx t$. Thus, it is a good approximation to take band and wavevector-independent tunneling matrix elements $|T_j^{ss'}({\bf k})| = \bar{T}$. Furthermore, in the typical case scenario the model parameters satisfy $\Gamma(\sim 10\,{\rm meV}) \ll\Delta (\sim 0.1-1\,{\rm eV})$. With this in mind, the integration of Eq.(\ref{eq:current}) allows an analytical expression to result in terms of line integrals in conical surfaces (see Fig.\ref{fig:resonance} and the discussion in Appendix A). Using $\mu_{1,2}=0$ and $V_{12} = \Phi_{12}$, we find that the non-resonant interlayer current takes the simple form
\be
\frac{I_{12}(x)}{I_{12}^{\rm (vl)}}=\frac{x^2-1/2}{\sqrt{2(x^2-1)}}, 
\quad
I_{12}^{\rm (vl)} = \frac{3\sqrt{2}\bar{T}^2}{4}\frac{G_{12}\Delta}{e}.
\label{eq:currentmin}
\ee
Here $x = eV_{12}/\Delta \gtrsim 1$ and $I_{12}^{\rm (vl)}$ is the valley current obtained at $x = \sqrt{3/2}$. When $eV_{12}/\Delta = 1$, however, the current is at resonance and reaches a maximum value which is sensitive to $\Gamma$. To leading order in $\Gamma$, we obtain (see Appendix B)
\be
 I_{12}^{\rm (pk)}/I_{12}^{\rm (vl)} = \pi \sqrt{\Delta/2\Gamma}, 
\label{eq:maxcurrent}
\ee
where $\Gamma$, in general, depends on the amount and type of disorder and/or temperature. Equations (\ref{eq:currentmin}) and (\ref{eq:maxcurrent}) yield $e \delta V/\Delta  \approx  3\sqrt{2} \bar{T}^2Z [\pi\sqrt{\Delta/2\Gamma}-1]/4 $. Importantly, very small values of $Z$ ($G_{23} \gg G_{12}$) make the bistable bias range negligibly small, whereas large values of $Z$ ($G_{23} \ll G_{12}$) would push the onset of the bistability region to very large bias potentials. Optimally, values of $Z \sim 1$ and very small $\Gamma$ would make the bistability effect more prominent. 

Achieving a large current ratio between bistable states is desirable for applications; this facilitates the reading process in a bistable device. From Eqs.(\ref{eq:currentmin}) and (\ref{eq:maxcurrent}), it is obtained that the current ratio between bistable branches is controlled by the parameter $Z\sqrt{\Delta / \Gamma}$. For realistic values of disorder, this ratio can be in the 1-20 ballpark. It is interesting to note that these already high values can be boosted by means of Pauli blocking. As shown in Figs.\ref{fig:resonance}c and d, for sufficiently heavily doped samples, the non-resonant bistable state (but not the resonant one) is Pauli-blocked. The degree of the electrical current ratio enhancement depends on second order processes which assist tunneling, such as scattering with defects or disorder. These second order processes are not considered here. 

The geometric control of $Z$, an appealing aspect of our system, can be understood from the Bardeen Transfer Hamiltonian Theory \cite{bardeen,wolf}. In this theory, the interlayer coupling $t$ is calculated from the overlap of the wavefunctions of layers $i$ and $j$ in the barrier region, $t = (\hbar^2/2m_e)\int d{\bf S}\cdot (\psi_i^*\nabla\psi_j - \psi_j\nabla\psi_i^*) $, with $d{\bf S}$ being a surface area element. Considering electrons tunneling across a square potential barrier with a height much larger than the electron kinetic energy, a tunneling matrix element of the form $t \propto {\rm exp}( - d_{ij} / \lambda)$ is obtained, where $\lambda$ is the WKB decay length defined above. The expression of $Z$ in Eq.(\ref{eq:parameters_2}) results from assuming barriers between layers 1 and 2 and between layers 2 and 3 are of the same material, in combination with Eq.(\ref{eq:conductance}). 

Although electrostatic doping of the graphene layers is not essential for the physics that we describe, it is a convenient feature of bistability. In particular, for a fixed external bias potential, each bistable state exhibits different carrier concentrations. Thus, any in-plane measurement, such as conductance or magneto-transport, will be able to distinguish two distinct bistable states. Indeed, from the inset of Fig.\ref{fig:current}c we see that the interlayer bias potential for each bistable state differs by an amount $\delta V_{12} \sim \Delta / e$ (see also discussion in the Appendix A). Taking into account the capacitance of the layers, the induced carrier difference between both states is approximately $\delta n \sim \kappa \Delta/4\pi e^2 d_{12}$ (here the quantum capacitance is not included). Using $\theta = 2^{\circ}$, $\kappa = 1$, and $d_{12} = 1\,{\rm nm} $, we obtain a carrier density difference $\delta n \sim 10^{12}\,{\rm cm^{-2}}$ between stable states. These large carrier density differentials can be used as a smoking gun of intrinsic bistability. 

\section{VI. OTHER GRAPHENE-BASED BISTABLE SYSTEMS}

Although we considered here for simplicity a two-step sequential tunneling structure where only one pair of layers can be resonant, similar ideas apply to more complex structures. Interesting examples include a two-step resonant-resonant structure, opening the possibility for tristability or multi-step ``cascade'' devices. 

Finally, we also expect bistable $I$-$V$ characteristics in twisted graphene trilayers in the absence of any dielectric material. Indeed, incommensurability between graphene lattices already suppresses interlayer hybridization, regardless of the layers being spatially separated by a fraction of a nanometer, thus enabling the sequential tunneling regime \cite{twistedmd1}. Furthermore, the massless Dirac spectrum, and thus Eq.(\ref{eq:H0}) and the subsequent transport model, remain valid but with a modified Fermi velocity \cite{twistedacn}. We stress, however, that stacked dielectric materials have two important advantages: (i) they enable tuning the interlayer coupling and (ii) they facilitate the interlayer potential build-up in order to achieve a resonant behavior. 

\section{VII. SUMMARY}

In summary, graphene-based van der Waals heterostructures afford a new platform to realize devices with tunable $I$-$V$ characteristics, in particular those with intrinsically bistable and hysteretic behavior. System parameters required to realize the bistable behavior are readily accessible in on-going experiments. The atomic scale interlayer distances can result in a fast response and large packing-densities, making these heterostructures appealing for a variety of applications. 

\section{ACKNOWLEDGEMENTS}

We thank A. D. Liao for useful discussions, and acknowledge support from National Science Foundation Grant No. DMR-1004147 [J.F.R.-N. and M.S.D.], from the STC Center for Integrated Quantum Materials, NSF Grant No. DMR-1231319, and from the Center for Excitonics, an Energy Frontier Research Center funded by the U.S. Department of Energy, Office of Science, Basic Energy Sciences under Award No. DE-SC0001088 [L.S.L.].

\appendix

\renewcommand{\thefigure}{A\arabic{figure}}
\renewcommand{\theequation}{A\arabic{equation}}
\setcounter{equation}{0}
\setcounter{figure}{0}

\section{Appendix A: Capacitance effects}

In the main text, we fixed the Fermi energy $\mu_i$ of the different graphene layers to some constant value. A more refined model of the $I$-$V$ response should, however, include quantum capacitance effects so that Fermi energy is allowed to vary with $V$. Although the features of bistability are not significantly modified by such corrections, as shown below, carrier density differentials between the bistable states are a smoking gun of intrinsic bistability. These electrostatic considerations are discussed next. 

\begin{figure}
\centering \includegraphics[scale=1.0]{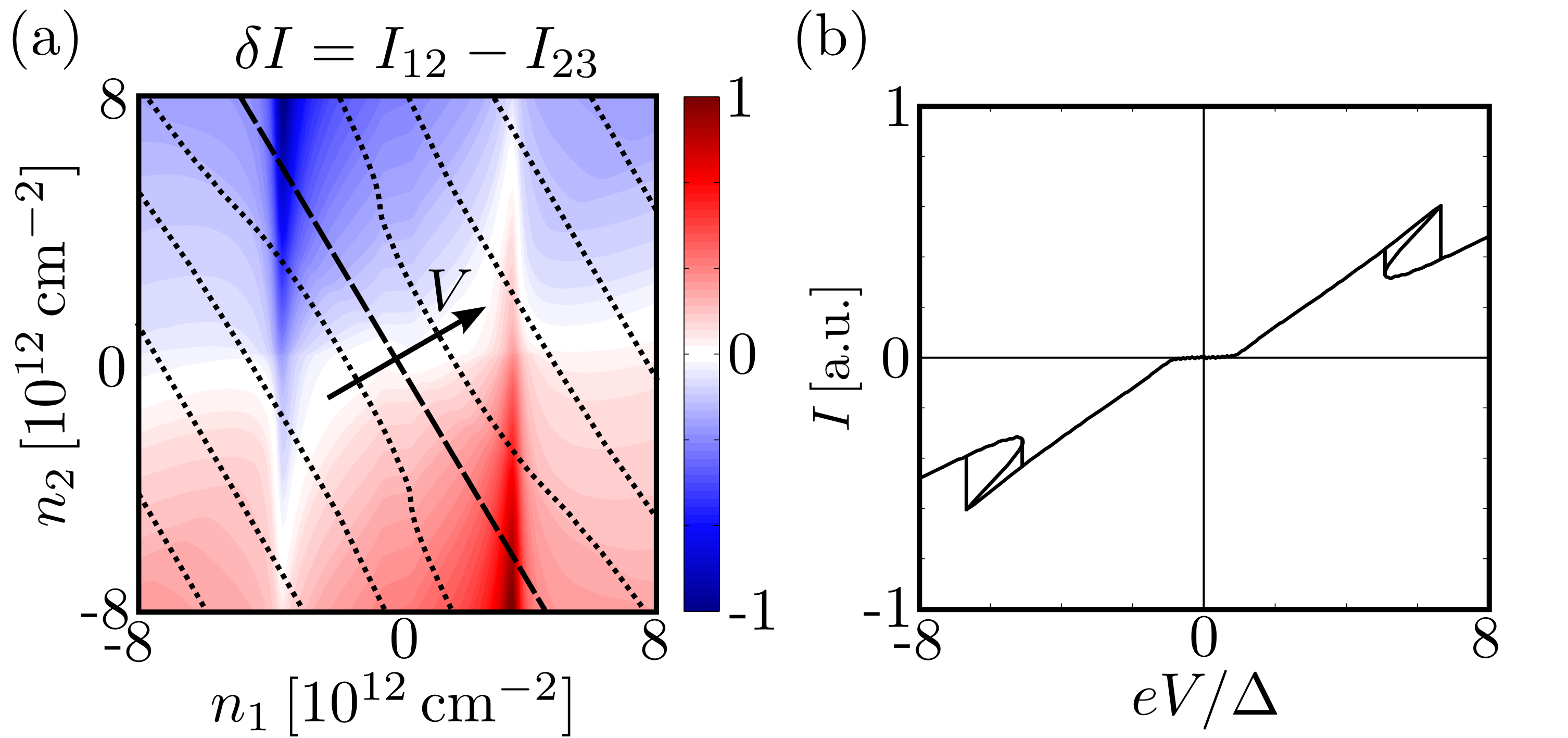}
\caption{Self-consistent bistable solutions including quantum capacitance effects [Eqs.(\ref{eq:cn})--(\ref{eq:biaseq})]. For fixed $V$, we find $n_1$ and $n_2$ such that $\delta I = I_{12} - I_{23} = 0$. The bias isolines from Eq.(\ref{eq:biaseq}) are marked with dashed ($V=0$) and dotted (finite $V$) lines, with an arrow pointing towards increasing $V$. The self-consistent $I$-$V$ curve, obtained from the intersection of $\delta I = 0$ and the $V$-isolines in (a), is plotted in panel (b).}
\label{fig:maps}
\vspace{-5mm}
\end{figure}

Here we numerically solve Eqs.(\ref{eq:current})-(\ref{eq:biaseq}), assuming a thin device separated by dielectric barriers of thickness $d_{12}= d_{23}=1.4\,{\rm nm}$ (e.g. four layers of hBN) and dielectric constant $\kappa = 5$. The procedure to solve the $I$-$V$ response self-consistently is shown in Fig.\ref{fig:maps}a, where $n_1$ and $n_2$ are taken as independent variables [$n_3$ is obtained from Eq.(\ref{eq:cn})], and $\delta I = I_{12}-I_{23}$ in Eq.(\ref{eq:current}) is numerically calculated (color map). For fixed $V$, indicated with dotted isolines in Fig.\ref{fig:maps}a, the self-consistent solutions to the equilibrium equations are given by the pair ($n_1$,$n_2$) such that $\delta I = 0$.

The resulting $I$-$V$ response is shown in Fig.\ref{fig:maps}b. Importantly, the $I$-$V$ characteristics are qualitatively similar to those obtained in the main text by neglecting quantum capacitance effects. Furthermore, by inspection of the $n_1$ and $n_2$ axes in Fig.\ref{fig:maps}a, we see that the difference in carrier concentration $\delta n$ between each bistable state is on the order of $\delta n \sim 10^{11}$-$10^{12}\,{\rm cm^{-2}}$. These carrier concentration differences can easily be detected by lateral transport measurements and may act as clear fingerprints of intrinsic bistability.

\section{Appendix B: Analytic expressions for the peak and valley resonant tunneling current}

We derive here Eqs.\,(\ref{eq:currentmin}) and (\ref{eq:maxcurrent}) of the main text, obtained under the assumption that the tunneling matrix elements $T_j^{ss'}$ in Eq.(\ref{eq:phasefactor}) are independent of the wavevector and band index, i.e., $|T_j^{ss'}(\mathbf{k})| = \bar{T}$. Under this assumption, $I_{12}$ depends only on the modulus of ${\bf q}_j$ but not on its direction, and $\sum_j |T_j^{ss'}(\mathbf{k})|^2 = 3\bar{T}^2$. Given that $\Gamma \ll \Delta$, when $e\Phi_{12}>\Delta$ (non-resonant state) we can set $\Gamma \rightarrow 0$ and thus take $A_{i,s}(\mathbf{k},\omega)=2\pi\delta (\omega - s\hbar v_{\rm F}|\mathbf{k}|)$. The two $\delta$-functions appearing in Eq.(\ref{eq:current}) can then be integrated in ${\bf k}$-$\omega$ space, resulting in a one-dimensional integral along the contour of an ellipse:
\be
\begin{array}{r}
\displaystyle
\sum_{ss'}\int\frac{d{\bf k}}{(2\pi)^2}\int_{\omega_1}^{\omega_2}\frac{d\omega}{2\pi} \delta(\omega - s\hbar v_{\rm F}|{\bf k}|)\delta(\tilde{\omega}-s'\hbar v_{\rm F}|{\bf k+q}|)\\
\displaystyle
 = \frac{\delta_{s,-}\delta_{s',+}}{16\pi^3 (\hbar v_{\rm F})^2}\int_{\phi_1}^{\phi_2}d\phi \frac{(e\Phi_{12})^2 - \Delta^2\sin^2\phi}{\sqrt{(e\Phi_{12})^2 - \Delta^2}}.
\end{array}
\label{eq:pwint}
\ee
Here we denote $\tilde{\omega}$ by $\tilde{\omega} = \omega+e\Phi_{12}$. In addition, the limits of integration on $\omega$ are given by $\omega_1 = e\Phi_{12}+\mu_2$ and $\omega_{2} = \mu_1$, whereas the limits of integration on $\phi$ are 
\be
\phi_i = \left\{ \begin{array}{lc} \pi/2,  & x_i > 1 \\ {\rm sin}^{-1}(x_i), & -1<x_i<1 \\ -\pi/2, &  x_i<-1 \end{array} \right. ,  \,\,\, x_{1,2} = \frac{2\mu_{1,2} \pm e\Phi_{12}}{\Delta}.
\label{eq:phii}
\ee
In obtaining Eq.(\ref{eq:pwint}), we parametrized ${\bf k}$-space using coordinates $k_x = k_r\sin \phi / 2$ and $k_y = \sqrt{k_r^2 - q^2}\cos \phi /2$, with ${\bf q}$ conveniently aligned in the $x$-direction. The integration over $k_r$ absorbs the first $\delta$ function, setting $k_r = e\Phi_{12}/\hbar v_{\rm F}$. Integration over $\omega$ absorbs the second $\delta$ function, fixing the limits of integration $\phi_{1,2}$ in Eq.(\ref{eq:phii}). Importantly, because $\Phi_{12}> \Delta $, the two $\delta$-functions in Eq.(\ref{eq:pwint}) can only be non-zero simultaneously when $s=-$ and $s'=+$ (i.e. holes of layer $1$ tunnel into electronic states of layer $2$, see Fig.\ref{fig:resonance}). Using $\mu_{1,2}=0$ and $V_{12} = \Phi_{12}$, Eqs.(\ref{eq:pwint}) and (\ref{eq:phii}) result in Eq.(\ref{eq:currentmin}). 

When $e\Phi_{12}=\Delta$, it is necessary to restore the finite linewidth to the Lorentzian spectral function $A_{i,s}(\mathbf{k},\omega)=2\Gamma/\left[(\omega-s\hbar v_{\rm F}|\mathbf{k}|)^2+\Gamma^2\right]$. In this case, the integral for the tunneling current yields
\be
\begin{array}{l}
\displaystyle
\sum_{ss'}\int\frac{d{\bf k}}{(2\pi)^2}\int_{\Phi_{12}+\mu_2}^{\mu_1}\frac{d\omega}{2\pi} A_{1,s}({\bf k},\omega) A_{2,s'}({\bf k + q},\omega)=\\
\displaystyle
= \frac{2}{(\hbar v_{\rm F})^2 \sqrt{\Gamma\Delta}} \left[ \int_{\mu_2}^{\Delta+\mu_1}d\omega \left|\omega(\omega - \Delta)\right|^{1/2}+{\cal O}(\Gamma/\Delta)\right].
\end{array}
\label{eq:maxint}
\ee
In obtaining Eq.(\ref{eq:maxint}), we transformed the integral of the spectral functions into a dimensionless integral of the form $I_{\rm res}(\epsilon) = \int d^2{\bf x}\{[f({\bf x})^2+\epsilon][g({\bf x})^2+\epsilon]\}^{-1}$. The functions $f$ and $g$ satisfy $f(0) = g(0) = 0$ and have a null Jacobian ${\rm det}[\partial_{\bf x}f,\partial_{\bf x}g](0) = 0$ (here $\epsilon = \Gamma / \Delta$). It can be shown that $I_{\rm res} \propto \epsilon^{-1/2}$, when $\epsilon \ll 1$. An expansion to leading order in powers of $\epsilon$ gives Eq.(\ref{eq:maxint}). Setting $\mu_{1,2} = 0$ in Eq.(\ref{eq:maxint}), the peak current $I_{12}^{\rm (pk)}$ in Eq.(\ref{eq:maxcurrent}) of the main text is obtained.


\begin{thebibliography}{10}

\bibitem{memoryreview} H.-S. P. Wong and S. Salahuddin, Nat Nanotechnol. {\bf 10}, 191 (2015).

\bibitem{flashmemory1}
A.~J. Hong, E.~B. Song, H.~S. Yu, M.~J. Allen, J. Kim,
  J.~D. Fowler, J.~K. Wassei, Y. Park, Y. Wang, J. Zou,
  R.~B. Kaner, B.~H. Weiller, and K.~L. Wang, ACS Nano {\bf 5}, 7812 (2011).

\bibitem{flashmemory2}
S. Bertolazzi, D. Krasnozhon, and A. Kis, ACS Nano {\bf 7}, 3246 (2013).

\bibitem{resistivememory1}
X. Wang, W. Xie, J. Du, C. Wang, N. Zhao, and J.-B. Xu, Adv. Mater. {\bf 24}, 2614 (2012).

\bibitem{resistivememory2}
Y.~J. Shin, J.~H. Kwon, G. Kalon, K.-T. Lam, C.~S.
  Bhatia, G. Liang, and H. Yang, Appl. Phys. Lett. {\bf 97}, 262105 (2010).

\bibitem{fetmemory}
Y. Zheng, G.-X. Ni, C.-T. Toh, M.-G. Zeng, S.-T. Chen, K. Yao, and B. {\"O}zyilmaz, Appl. Phys. Lett. {\bf 94}, 163505 (2009).

\bibitem{memristor0} D. B. Strukov, G. S. Snider, D. R. Stewart and R. S. Williams, Nature {\bf 453}, 80 (2008).

\bibitem{memristorreview}  J. J. Yang, D. B. Strukov and D. R. Stewart, Nat Nanotechnol {\bf 8}, 13 (2013).

\bibitem{magnetic} A. D. Kent, and D. Worledge, Nat Nanotechnol {\bf 10}, 187 (2015).

\bibitem{vdwheterostructures}
A.~K. Geim and I.~V. Grigorieva, Nature {\bf 499}, 419 (2013).

\bibitem{twistnovoselov}
A.~Mishchenko, J.~S. Tu, Y.~Cao, R.~V. Gorbachev, J.~R. Wallbank, M.~T., Greenaway, V.~E. Morozov, S.~V. Morozov, M.~J. Zhu, S.~L. Wong, F.~Withers, Y-J. Woods, C. R.~Kim, K.~Watanabe, T.~Taniguchi, E.~E. Vdovin, O.~Makarovsky, T.~M. Fromhold, V.~I. Fal'ko, A.~K. Geim, L.~Eaves, and K.~S. Novoselov, Nature Nano {\bf 9}, 808 (2014).

\bibitem{resonanttunneling}
L.~Britnell, R.~V. Gorbachev, A.~K. Geim, L.~A. Ponomarenko, A.~Mishchenko, M.~T. Greenaway, T.~M. Fromhold, K.~S. Novoselov, and L.~Eaves, Nature Commun. {\bf 4}, 1794 (2013).

\bibitem{goldman}
V.~J. Goldman, D.~C. Tsui, and J.~E. Cunningham, Phys. Rev. Lett. {\bf 58}, 1256 (1987).

\bibitem{sheardtombs}
F.~W. Sheard and G.~A. Toombs, Appl. Phys. Lett. {\bf 52}, 1228 (1988).

\bibitem{goldman2}
A.~Zaslavsky, V.~J. Goldman, D.~C. Tsui, and J.~E. Cunningham, Appl. Phys. Lett. {\bf 53}, 1408 (1988).

\bibitem{relaxation}
P.~A. George, J. Strait, J. Dawlaty, S. Shivaraman, M. Chandrashekhar, F. Rana and M.~G. Spencer, Nano Lett. {\bf 8}, 4248 (2008).

\bibitem{twistedacn}
J.~M.~B. Lopes~dos Santos, N.~M.~R. Peres, and A.~H. Castro~Neto, Phys. Rev. Lett. {\bf 99}, 256802 (2007).

\bibitem{twistedmd1}
R.~Bistritzer and A.~H. MacDonald, Phys. Rev. B {\bf 81}, 245412 (2010).

\bibitem{twistedmd2}
R.~Bistritzer and A.~H. MacDonald, PNAS {\bf 108}, 12233 (2011).

\bibitem{brey} L. Brey, Phys. Rev. Appl. {\bf 2}, 014003 (2014).

\bibitem{qc}
S.~Luryi, Appl. Phys. Lett. {\bf 52}, 501 (1988).

\bibitem{bardeen}
J. Bardeen, Phys. Rev. Lett. {\bf 6}, 57 (1961).

\bibitem{wolf}
E. L. Wolf, {\it Principles of Electron Tunneling Spectroscopy}, Second Edition (Oxford University Press, Oxford, 2012)

\end{thebibliography}
\end{document}